# Modelling of LIDAR sensor disturbances by solid airborne particles

M. Hadj-Bachir[1], P. de Souza[1], P. Nordqvist[2], N. Roy[2]

1: Virtual System & Control ESI-Group
2: Volvo Autonomous Solution

**Abstract**: This paper aims to introduce a method for simulating with a real time performance the automotive LIDAR disturbance by dust clouds caused by natural phenomena, mechanical or man-made processes like a traveling vehicle. In this study, we are interested to study the interaction of an automotive LIDAR sensor with a dust cloud composed of solid particles. The main objective of this study is to provide a simulation model to industry and research laboratories that help to study LIDAR performance in a dust-sand environment with the capability to reproduce the encountered problems in degraded conditions and the ability to parameterize the degradation model. Based on industrial projects with a passenger's vehicles and truck manufacturers, we present LIDAR sensor and functionalities to perceive objects in a scene (pedestrian, car, truck, ...) in clear or extreme weather conditions. Simulated and experimental data are compared and analyzed in this article. The features presented are evaluated according to their quality for object detection. This study can be applied to sensors post-processing algorithms (object recognition, tracking, data fusion...) and even to the design of cleaning systems.

## 1. Introduction

For autonomous driving to be safe and reliable, perception functions must be supported by sufficient artificial vision. Accounting for the sensors' performance has become very important to ensure their reliability and therefore ensure the safety of the vehicle even under adverse environmental influences like rain, dust, snow [1] [2] or by contamination on the front of the sensor screen [3]. With the vast selection of available sensors and seemingly endless options in terms of range, resolution, or sensitivity, it is difficult to select the optimal system for one's Autonomous Vehicle (AV) application [4] [5]. LIDAR sensors have become very popular in the robotics and automotive domain due to their ability to provide high-density data with accurate range measurements while being relatively robust to lighting conditions. LIDAR sensors capture high definition data from the environment as point clouds. Such a sensor makes use of multiple scanning layers in order to be able to detect obstacles at different heights even when the pitch of the vehicle changes. This renders LIDAR a premium choice for a perception system on a vehicle [6] [7]. However, as advanced as LIDAR technology may be today, no current single sensing technology can guarantee a perfect accuracy of the measurement. Indeed, LIDAR sensors have several drawbacks to consider [18]. Low vertical resolution: in low cost LIDAR models, which have fewer layers [9]. Poor detection of dark and specular objects: a black target with a non-Lambertian material that does not scatter radiation back to receiver can appear as invisible to the LIDAR, since the most radiation is absorbed by the target. Affected by weather conditions: adverse weather conditions are significantly impacting the performance of LIDAR based scene understanding by causing undesired measurement points that in turn effect missing detections and false positives [10]. In harsh condition [11] [12], dense airborne particles (water, fog, smoke, dust...) could be misinterpreted as an obstacle ahead of the vehicle, which could possibly result in an emergency braking of the AV.

Studies of sand and dust cloud effects remain immature for automotive LIDAR performance compared to RADAR. This delay is mainly due to: the higher costs of LIDAR devices compared to RADAR; less LIDARs being using in automated driving systems compared to RADAR and camera; Due to eye safety and the OEM requirements in terms of detection ranges, suppliers developpe LIDARs ranging from 900 nm to 1.5 micrometers wavelength; Based on optical signal transmission and reception, laser propagation is influenced by weather phenomena as demonstrated in [1, 11, 12, 13].
Because of its influence on light propagation, dust is an issue for sensing by LIDAR. It is to be noted that LIDAR sensor manufacturers already make efforts to mitigate dust effects by adding specific filtering strategies to the sensors, either through software post-processing of the detection output or by optical filtering of the return light pulse. In [14] Lafrique P & al have modeled an atmospheric multiwavelength LIDAR in a dust aerosols layer, where optical properties are calculated for different shapes and with or without an overlay of water. Two different shapes of dust particles are simulated, spheres and parallelepipeds with a coating or not. Results of this study show significant effects on multiwavelength LIDAR signals. The potential gain of this study is the using LIDAR wavelengths in the infrared spectral domains to access on more reliable microphysical properties of the aerosol. TG Phillips & al classify four effects of airborne particles on LIDAR sensors [12]. The authors had tested three LIDAR sensors, and they found that all sensors exhibit the same behaviors under similar test conditions. The LIDAR sensitivity to airborne particles such as dust or fog can lead to perception algorithm failures e.g. the detection of false obstacles by autonomous vehicle. Leo Stanislas & al have addressed this problem by proposing methods based on deep learning approaches to classify airborne particles in LIDAR data point [10].

As a summary, we have identified publications in the literature that deal with data analysis of physical experiments, and with noise filtering algorithms generated by dust cloud. In our study, we are analyzing the behavior of a simulation model used in a virtual testing tool, that can predict the detection performance and output in extreme weather conditions such as fog, snow, rain, smoke and dust. The modeling of sensor disturbances must answer two objectives:

- consider the physics to obtain high quality of data.
- the simulations must be fast enough to compute so that many scenarios can be evaluated in a reasonable amount of time.

Our study aims to answer to these two objectives. In the rest of the document, we will first briefly describe the LIDAR disturbance by dust clouds experiments and we will summarize the major results of these experiments. Then we will show simulation cases of low and high density of particles dust cloud and conclude.

## 2. Experimental results

The objective of the experiments carried out in this study is to observe the effect of different dust cloud parameters on the automotive LIDAR. However, dust present in the air can have many possible shapes, sizes and densities. Characterizing dust clouds adequately can be complex due to the high variability. In this study, laboratory conditions are chosen over open world conditions to minimize the variance and the unknown factors.

Experimental campaigns are carried out where two automotive LIDAR sensors disturbed by different dust clouds are tested. The tested LIDAR are OS-1 from Ouster [15] and VLP-16 from Velodyne [16] company. Different variants of dust cloud volumes, densities as well as composition (flour, cement, calcium carbonate) were tested. Moreover, the influence of the separation distance between the cloud and sensor is tested.

Standard video cameras are used to record the dimensions and the evolution of the dust cloud. One camera faces the front of the dust cloud (regarding the location of the lidar sensor), and the second is located on the left side. The experiment was carried out in garage where a car is placed at 16m from the LIDARs and a truck at 40 m (figure 1a).

Figure 1b represents the reference image of the Ouster OS-1 LIDAR. This image is used for comparison when the sensor is disturbed by cloud dust. In the LIDAR images, the detection returns corresponding to the car, the truck and the experimenter have been circled. From the analysis of previously published experimental data, it is expected that particles of dust appear as a cloud point in LIDAR images. This returned light is caused by the backscattering phenomenon. In addition to this backscattering effect, the intensity (reflectivity), as well as the number of returned points from targets (car, truck) is also attenuated by the dust clouds. These behaviors are observed for the Ouster OS-1 and Velodyne VLP 16.

The experiment results shown in figures 1c and 1d correspond to the effect of the dust cloud. The dust cloud is generated at 6 m from the sensor. In figure 1c, the LIDAR is disturbed by a low dust density, while the high dust density is represented in figure 1d. When comparing the figure 1b (reference), and 1c (low density) we can see that the noise (backscattering light) generated by the dust particles is low and the car and truck remains detectable. If we compare figure 1b (reference) and 1d (strong density), we can observe that the dust cloud obscures completely the car and the truck. The intensity (reflectivity) of the particle backscattering light is more intense compared to the case of low dust cloud density. This backscattered light then causes the LIDAR glare. The other results of the different experiments variants showed that:

- the backscattering light is intense in the front side compared to the back side of the dust cloud volume.
- The backscattering light from dust cloud decreases as the distance between LIDAR and dust cloud increase.
- All types of mater used to create the dust cloud (cement, floor, calcium, ....) reflect the backscattered light.

## 3. Simulation results

We have developed a model to simulate LIDAR sensors disturbed by airborne particles like smoke and dust, and the model was named ALDUS (Automotive LIDAR DUSt). The model estimates in a computational cost-efficient way the energy attenuation as well as the intensity of the backscattered light issued from airborne particles of the dust or smoke cloud.

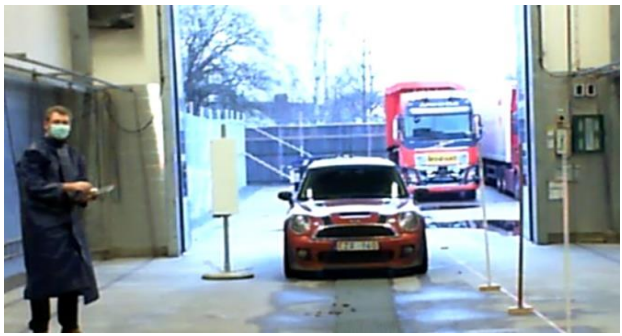
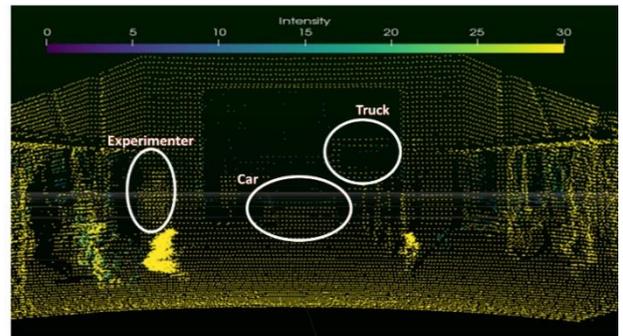
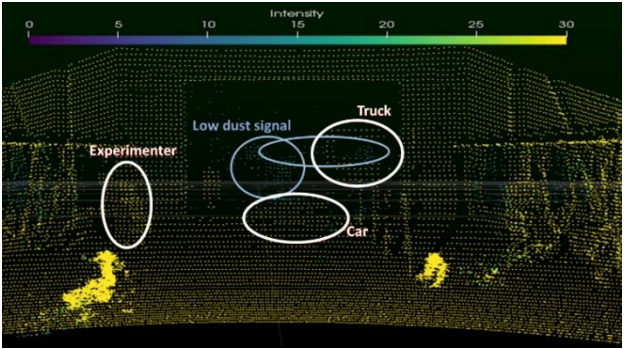
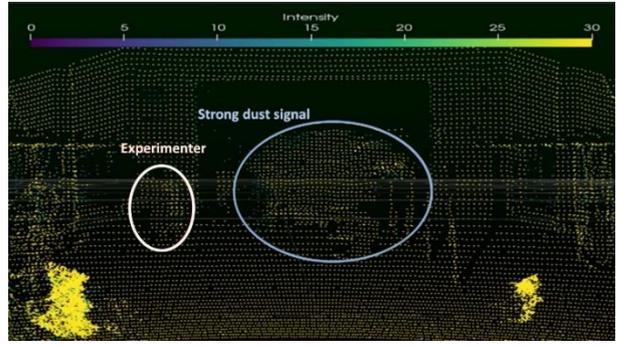

Figure 1: experiment environment and Ouster OS-1 LIDAR data. a) environment, b) LIDAR reference image, c) low dust cloud density disturbance, d) strong dust density disturbance

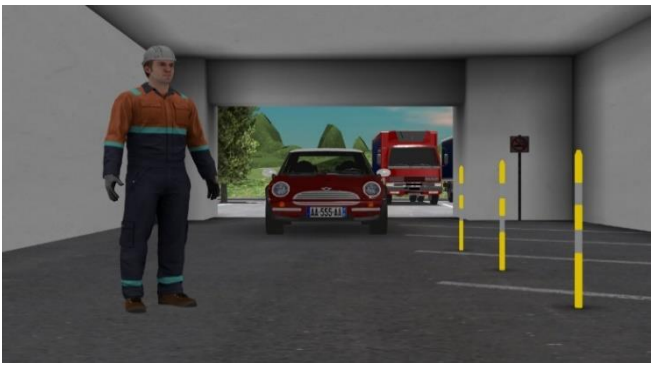
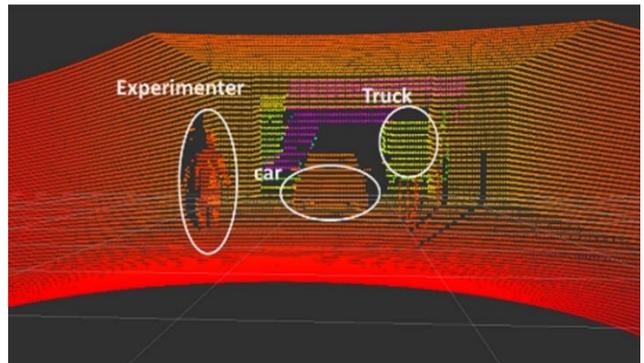
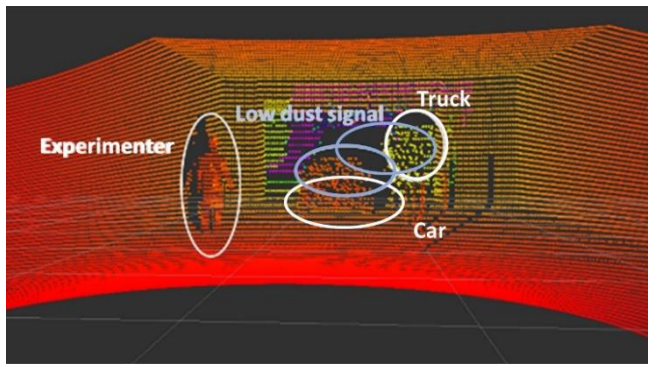
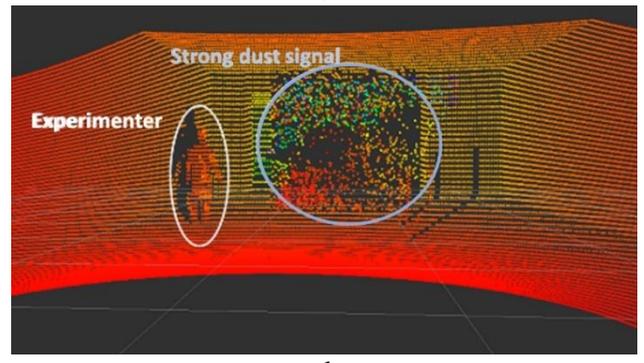

Figure 2: Simulation of experiment environment and simulation of lidar data in PROSIVIC. a) environment, b) lidar reference image, c) low dust cloud density disturbance, d) strong dust density disturbance

The model is implemented in the PROSIVIC simulation platform and can also operate as a standalone model executable to simulate lidar output affected by dusty conditions. The connection of the ALDUS model interfaces to other simulation platforms can be either dynamic (FMI, DDS standards) as co-simulation, or static to process recorded data.

In this first version, the ALDUS model for the dust considers spherical and homogeneous particles where the size and density of airborne particles as well as the volume of the dust/smoke clouds can be configured. The dust is assumed to occupy only a portion of space, and with a configurable shape. This simulation model helps to study LIDAR performance in a dust-sand, smoke environment with the capability to reproduce the encountered problems in degraded conditions and the ability to parameterize the degradation model.

Regarding the validation of the ALDUS model, we used the software simulation environment PROSIVIC to reproduce the experiment described in section 2. The scene 3D model was modelled based on the dimensions of the real scene, but no 3D scan-based modelling was performed. The simulation environment is illustrated in image 2a where the car and truck are placed at 16 m and 40 m respectively from the LIDAR position. Note that image 2a is the equivalent of image 1a with some small differences in the number of objects such as boxes and electric cables. However, the virtual scene (Figure 2a) could be improved by adding the objects or by the import of real 3D scans, yet this is not needed for our objectives. In this study, the details in terms of the scene objects numbers and rendering are not important, because we focus only on the detection of the car and truck by the LIDAR with dust cloud disturbance. The parameters to simulate the LIDAR characteristics like the optical performance, laser technology, and LIDAR output are considered and implemented in PROSIVIC. These parameters are taken from the sensor datasheet [15] and [16].

Figures. 2b, 2c and 2d correspond respectively to the simulation of LIDAR sensor without disturbance, disturbance with a low dust particles density, and disturbance with strong dust particles density.

The comparison of figures 1b and 1c, shows the capacity of PROSIVIC to simulate the Ouster LIDAR considering the laser parameters, the field of view, resolution and delivers output data in a format identical to a real LIDAR (coordinates of cloud points, distance, angles, reflectivity). In addition, the LIDAR model can easily simulate different scanning technologies like mechanical LIDAR, solid-state, and flash LIDARs. When comparing respectively figures 1c and 2c or 1d and 2d, we can see that the experimental result and simulation are in good agreement.

## 4. Benefits of using sensors physics simulation

Volvo Autonomous Solutions is interested in dust simulation because it matches closely with the current use case where we conduct pilot projects in confined areas. In particular, Volvo Autonomous Solutions are involved in mines operations, where dust will be present in high volumes. Simulating this will help us take mitigating actions before deploying the solution on the site. A high-fidelity simulation model will support both the safety and the production goals. The results shown in section 3 confirm the ability to reproduce virtually sensors disturbance experiences. In addition, the parameters characterizing the dust cloud are fully controlled in simulation which is difficult to obtain during the experimental phase. Moreover, several critical situations in road environments, deserts, or on construction sites were simulated to test the sensors robustness in extreme conditions. This allowed us to determine the dust clouds types and density which can generate noise or sensor failure. Usually real testing has several drawbacks such as difficulty in reproducing test conditions, but also costs and risks associated with using a real car in an outdoor, and sometimes open, environment. On the other hand, the costs of integration problems are very high when they are detected too late in the development cycle. Additionally, in order to validate each new system, the testing process requires traveling thousands of miles in many types of scenarios and environments, with a limited warranty that all conditions are statistically met [17]. The use of simulation from the start of the development process of a system or sub-systems, therefore, allows manufacturers and scientists to anticipate malfunctions and complete validation tests. Depending on the scope of each system or sub-system to be validated and depending on the level of representativeness of the simulations, the way of simulating detection systems may be different. Ranging from fully idealized sensing function to detailed physical simulation of the sensor, these models offer a different value in the evaluation process compared to actual testing. Yet with the growing ambition to produce high-fidelity virtual prototypes of autonomous vehicles, the physics-based simulation sensor is becoming a necessity to address system performance in operation.

## 4. Conclusion

In this paper, we have studied the effect of airborne particles on LIDAR detection. The experimental data shows that the LIDAR sensor collects a backscatter light where the signal intensity is subject to the dust configuration and position. Based on the experimental data analyses and literature, we had developed a disturbance model which helps simulate LIDAR disturbed by airborne particles cloud. The actual model considers the cloud volume, particles size, and density. This disturbance model helps us to simulate the target reflectivity and laser energy attenuation under the airborne clouds. In addition, the model provides synthetic noise data of created by dust/sand cloud which can be injected into post-processing or artificial intelligence algorithms. This study represents a foundation for dust simulation, by showing how its fundamental effects can be simulated while being computationally efficient. Further improvements can aim to simulate actual real-world conditions.

As a possible improvement of the simulation model, the modelling of the structure of the dust cloud can be considered. The inhomogeneity of the airborne dust cloud including cloud 3D geometry, different particles species, particles geometry – is expected to influence LIDAR detection. Indeed, the different particles species absorb and reflect laser energy in different manner. Also, simulating the evolution in time of a dust cloud could help better address operational driving scenarios. By using a generator of dynamic particle model, the effect of temperature, pressure can be accounted used to better reproduce a real dust cloud. This particle generator will estimate the exact particles density and particles projection in 3D space. The application of this model will help for example to simulate the effect of dust lifted by vehicles' tires, which is a common behavior of dust.

## 5. Acknowledgements

The authors of this paper wish to thank and acknowledge the support from Vinnova, the Swedish Governmental Agency for Innovation Systems.